# A digital microarray using interferometric detection of plasmonic nanorod labels

Derin Sevenler[1], George Daaboul[2], Fulya Ekiz-Kanik[1], and M. Selim Ünlü[1]

1. Department of Electrical and Computer Engineering, Boston University, Boston, MA
2. NanoView Diagnostics, Boston, MA

**PRELIMINARY DRAFT**
**23 January 2017**

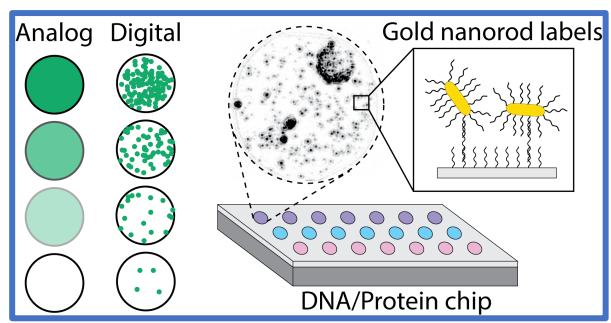

**DNA and protein microarrays are a high-throughput technology that allow the simultaneous quantification of tens of thousands of different biomolecular species. The mediocre sensitivity and dynamic range of traditional fluorescence microarrays compared to other techniques have been the technology's Achilles' Heel, and prevented their adoption for many biomedical and clinical diagnostic applications. Previous work to enhance the sensitivity of microarray readout to the single-molecule ('digital') regime have either required signal amplifying chemistry or sacrificed throughput, nixing the platform's primary advantages. Here, we report the development of a digital microarray which extends both the sensitivity and dynamic range of microarrays by about three orders of magnitude. This technique uses functionalized gold nanorods as single-molecule labels and an interferometric scanner which can rapidly enumerate individual nanorods by imaging them with a 10x objective lens. This approach does not require any chemical enhancement such as silver deposition, and scans arrays with a throughput similar to commercial fluorescence devices. By combining single-nanoparticle enumeration and ensemble measurements of spots when the particles are very dense, this system achieves a dynamic range of about one million directly from a single scan.**

Protein and DNA microarray technologies continue to be useful in a myriad of biomedical and clinical applications, such high-throughput genetic or transcriptional analysis and multiplexed protein detection. Insufficient sensitivity and dynamic range are the two most commonly cited weaknesses of the technology, and have motivated the widespread adoption of newer methods based on DNA sequencing or sample compartmentalization to perform sensitive and multiplexed nucleic acid or protein analysis[1–3].

This practical limit is not imposed by the microarray format itself, but rather the sensitivity of conventional fluorescence readers. A theoretically ideal transducer that could quantify the absolute number of immobilized targets with no background would be limited only by Poisson process variability[4]. Since a typical 100 μm-wide microarray spot contains about one billion probe oligonucleotides[5], the ideal microarray transducer would also have a dynamic range as high as 100 million.

In practice, virtually all fluorescent scanners have a dynamic range of only 100-1,000, despite the fact that single fluorophores are routinely detected in scientific microscopy[6–8]. The reason for this discrepancy is that single fluorophore detection requires a high numerical aperture (NA) objective lens with a tight focus tolerance and a small field of view, while microarrays are often larger than 1 cm$^2$. It is a significant technical challenge to maintain the required focus tolerance (usually less than 300 nm) while scanning across a large array. Most single-fluorophore scanners simply cannot scan large arrays, or otherwise require a sophisticated focus-tracking system[9–12]. Even then, the scanning throughput remains fundamentally limited by the quantum "speed limit" of the fluorophore's emission lifetime, which sets the fluorophore's minimum obtainable average time between photon emission events.

In contrast, measurements of light scattering by nanoparticles have no saturated emission rate or photobleaching. The speed and throughput of light scattering measurements therefore tend to be limited only by the available light power, or maximum allowable local heating of the particle. Gold nanoparticles are routinely used place of fluorescent probes for microarray labeling, and either detected individually directly based on their light scattering[13–21] or indirectly *via* silver deposition[22–25]. All of these techniques have successfully enhanced the sensitivity of microarrays by several orders of magnitude, and typically have a limit of detection of roughly 1 femtomolar. However, the former of these approaches have all required a high-NA lens—reducing throughput to less than 20 spots—while the latter methods suffer from reduced dynamic range unless multiple rounds of silver enhancement and re-scanning is performed.

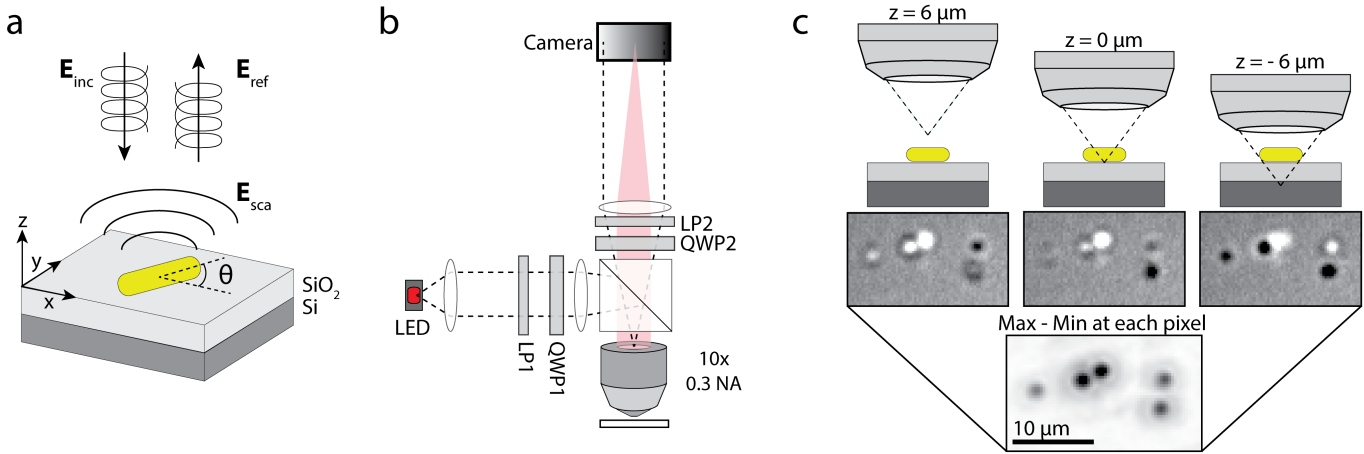

Figure 1. (a) Schematic of gold nanorod (GNR) detection with interferometric reflectance imaging sensing (IRIS). Circularly polarized plane wave illumination $E_{inc}$ is reflected as a circularly polarized plane wave by the IRIS substrate $E_{ref}$ but scattered by the GNR as a spherical wave $E_{sca}$ that is linearly polarized along the rod's longitudinal axis $\theta$. (b) Schematic of the reflectance microscope used to image the IRIS chip. Both the reflected (dotted lines) and scattered light (red shadow) are imaged onto the camera with a 10x objective (**LP**- linear polarizer, **QWP**- quarter wave plate). (c) At the camera, the phase between scattered and reflected light depends on both particle orientation $\theta$ and focus position $z$. All GNRs are robustly detected regardless of their orientation by acquiring a z-stack and measuring the difference between the maximum and minimum at each (x, y) pixel.

To our knowledge, no method exists to enumerate individual nanoparticle labels across a very large surface with a throughput comparable to commercial fluorescence scanners. The most obvious way to increase throughput would be to use a lower magnification lens to increase the instrument field of view. However, lower magnification lenses are less efficient at collecting light, and reduced light collection is unacceptable for many of these designs. For dark-field detection, the signal scales (approximately, for NA<0.6) with the fourth power of the NA.

Interferometric detection is more resilient than dark-field to very weak signals since the scattered light amplitude is measured rather than intensity. Interferometric reflectance imaging sensing (IRIS) is one of a family of similar optical techniques for interferometric detection of nanoparticles immobilized on a substrate[17,26,27]. IRIS uses a substrate of polished silicon with a thin film of thermally-grown silicon dioxide. The substrate is imaged with a reflectance microscope with Köhler illumination from an LED source. In the absence of any nanoparticle, the microscope camera observes a featureless reflection of the illumination light on the substrate surface. If a nanoparticle is present, light scattered by the particle is also imaged onto the camera where it forms a faint diffraction-limited interference pattern with the reflected light. The 'normalized intensity' of this interference pattern is obtained by dividing by the intensity of the reflected field alone.

If the substrate and illumination were both ideally smooth and uniform, arbitrarily weak signals could be detected by collecting enough photons until the shot noise was reduced below the normalized intensity of the signal. In practice, IRIS substrates are slightly heterogeneous, resulting in about 0.5% fluctuations in reflectivity across the chip surface. Therefore, the normalized intensity of nanoparticles must be at least 2-3%, to be robustly detected.

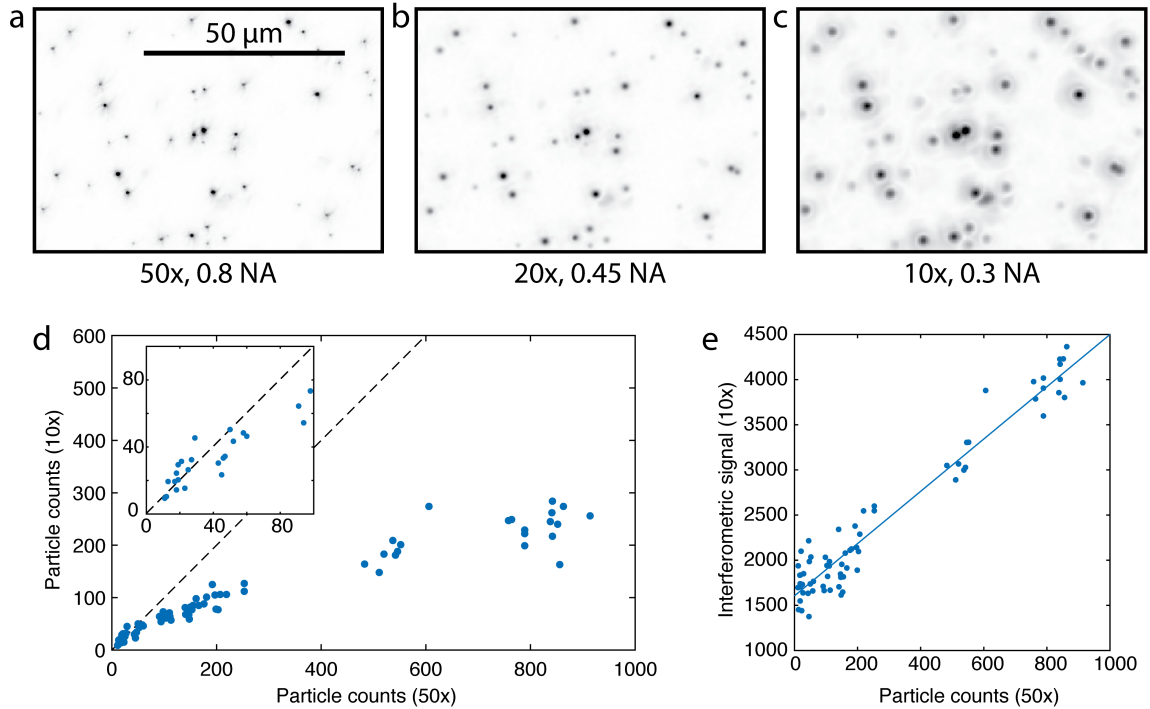

**Figure 2.** Validation of GNR quantification. Gold nanorods on an IRIS substrate imaged with (a) 50x, (b) 20x, and (c) 10x objective lenses. Individual nanorods are visible as diffraction-limited black spots. The contrast of the interferometric image has been increased by 1.5-fold and 3-fold in (b) and (c) respectively. Provided they are not too close to one another, all nanorods visible in (a) are also visible in (b) and (c). (d) Scatter plot of the number of particles counted within 70 different spots (cropped to 1e4 μm$^2$), with 10x vs. 50x objective lenses. While the two magnifications are in good agreement when the number of GNRs is low (inset), the lower resolution of the 10x objective results in systematic under-counting when there are over 60 particles per spot. (e) Scatter plot of interferometric signal of the same 70 spots versus number of particles counted with a 50x lens. The interferometric signal is proportional to GNR number when there are over 200 particles.

This is the main challenge to performing interferometric detection with a low-NA objective—reducing the NA reduces the collection of scattered light but not that of reflected light, which reduces the normalized intensity of the particle image. If the normalized intensity drops below the 3% visibility threshold, the particle will no longer be detectable with IRIS. The reduced scattering could be compensated for, if the reflected light could be attenuated. However, this cannot be done with a simple neutral density filter since it takes the same optical path from the chip surface to the camera. We therefore developed a method that utilizes the unique depolarization properties of plasmonic gold nanorods to selectively attenuate the reflected light.

**RESULTS AND DISCUSSION**

**Interferometric detection of single plasmonic nanorods with a 10x objective lens.** Plasmonic gold nanorods (GNRs) are rod-shaped nanoparticles with the interesting optical property that their scattering cross section is a function of both wavelength and polarization. At its longitudinal surface plasmon resonance wavelength, a GNR effectively only scatters the component of the excitation that is polarized along its longitudinal axis[28].

Consider a nanorod on an IRIS substrate illuminated by circularly polarized plane wave at normal incidence ($E_{inc}$, Figure 1a). Light reflected by the substrate ($E_{ref}$) remains circularly polarized, but the light scattered by the particle ($E_{sca}$) will be linearly polarized. This discrepancy between the polarizations of the scattered and reflected light can be exploited to selectively attenuate the reflected light with a quarter wave plate and linear polarizer in the imaging path (QWP2 and LP2, Figure 1b). These two optics are adjusted to attenuate the reflected light amplitude by precisely 95%, while attenuating the scattered light amplitude by only 30 to 50%, depending on polarization direction. This has the effect of increasing the normalized intensity of the GNRs by about 7 to 10-fold, depending on the particle's surface orientation angle $\theta$ (Figure S1). These optics also retard the phase of the scattered light by different amounts, depending on $\theta$. This variable phase shift causes particles of different orientations to appear darker than the background, or brighter, or nearly invisible when the two fields are in quadrature (Figure 1c). Fortunately, changing the focus position also changes the path length difference between the scattered and reflected light. All particles are made visible by taking a z-stack of images, and then subtracting the minimum value from the maximum value at each $(x, y)$ pixel location (Figure 1c). This mechanism is described in detail elsewhere[27], but it essentially requires that the illumination is aligned so as to radically under-fill the back pupil of the objective, and approximate a plane wave at the chip surface.

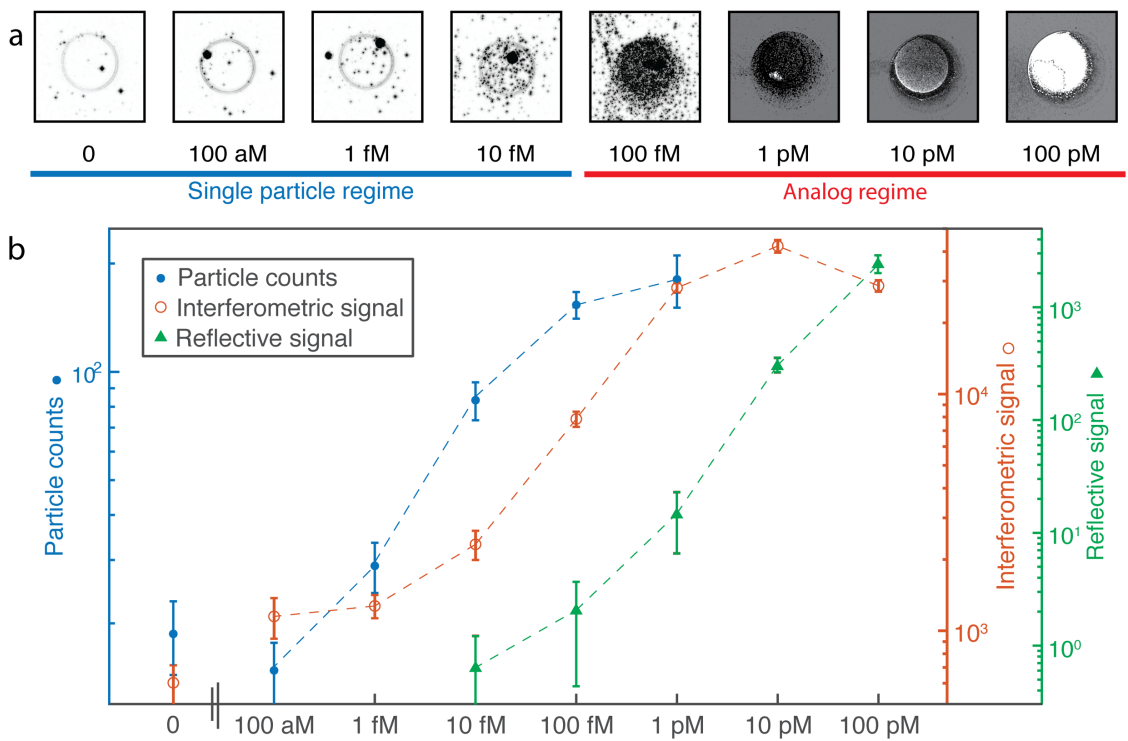

**Figure 3.** (a) DNA microarray spots on eight different IRIS substrates after incubation for 4 hours with GNRs conjugated with the complementary sequence at a range of concentrations. (b) Log-log standard curves of the number of individual particles counted, average interferometric signal and reflective signal per spot (described in text). Error bars indicate one standard deviation (n=10 spots per chip).

**Co-optimization of the optical system for rapid GNR detection.** The oxide film thickness of the substrate, illumination wavelength and nanorod geometry all effect the amplitude of the scattered light collected by the objective. We used a quantitative model of interferometric reflectance imaging described previously to co-optimize these various parameters and guide the selection of 25 nm diameter GNRs with a longitudinal surface plasmon resonance wavelength of 650 nm, a substrate oxide thickness of 110 nm, and 650 nm LED illumination source with an 10 nm FWHM bandpass filter (Figure S2)[27]. Early on in this study, we also compared the circular polarization scheme described here with a simpler cross polarization scheme, in which the illumination is linearly polarized rather than circularly polarized, and mostly blocked by a crossed polarizer in the collection path (Figure S3). However, we found that only particles with certain surface orientations were made more visible using the cross polarization scheme, while the circular polarization scheme enhanced the visibility of rods of all orientations.

Once the optical design was finalized, we experimentally determined the optimal amount of attenuation for the rapid detection of GNRs. We initially observed that the noise floor of the image began to increase noticeably when the analyzer (LP2, Figure 1b) was adjusted to attenuate over 99% of the reflected light. Attenuating the reference increases the normalized intensity of the particle, but also lowers the intensity of the collected light, which necessitates a longer exposure time to collect the same number of photons. In our apparatus, back reflections from the cube beam splitter were the main source of stray light and had a relative intensity of about 1% of the illumination (Figure S4). We found that setting the analyzer to extinguish precisely 95% of the reflected light was enough to enhance the normalized intensity the GNRs until they were clearly distinguishable against the background, without substantially increasing the noise floor.

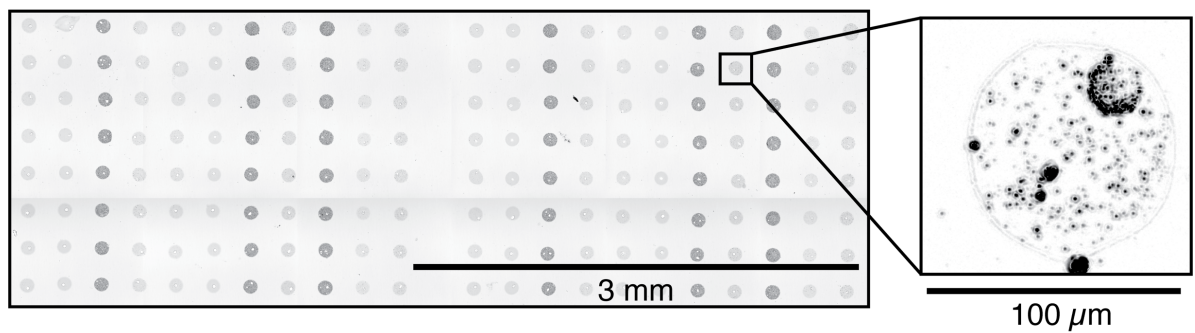

**Fig. 4.** A composite image of an IRIS microarray scanned by the automated digital microarray instrument. The instrument has a scan speed of about 3 mm$^2$ (50-125 spots) per minute, similar to most commercial fluorescence scanners.

**Validation of single GNR counting with a 10x lens.** To experimentally measure the visibility of GNRs with large field-of-view objectives, 25 nm by 71 nm rods were sparsely immobilized onto an IRIS substrate, which was dried and imaged with 10x, 20x and 50x objective lenses (Figure 2a-c). All individual nanorods that were visible in the 50x frame were also easily visible in those frames taken by the 20x and 10x objective lenses, provided the particles were far enough apart from one another to be distinguished.

We anticipated that the low magnification system would be able to accurately count nanoparticles when they were sparse, but would under-count them as the nanoparticle surface density increased until they were closer than the resolution limit. To investigate this, we printed microarrays of poly-adenine (14-A) single stranded DNA oligonucleotides on IRIS substrates, and conjugated gold nanorods with poly-thymine (14-T) single stranded DNA. The conjugated GNRs were diluted to a range of concentrations between 100 attomolar and 100 picomolar, then incubated with the DNA chips for 4 hours, then finally washed and dried. A total of 70 spots were imaged with instrument using both the 10x and 50x objective, and images were analyzed using custom particle analysis software described previously[29]. As anticipated, the 10x objective system accurately enumerated immobilized GNRs when they were few (i.e., 60 GNR per spot of fewer), but systematically under-counted as their number increased (Figure 2d). This was recapitulated by a simple model of particle crowding, in which GNRs in the image are modeled as disks placed randomly in an image with uniform probability, and any over-lapping disks are 'detected' as a single particle. In the model, accuracy was improved by measuring and compensating for the average rate of under-counting due to particle crowding (Figure S5). Although this is a statistical method, such an approach may be useful at improving the absolute accuracy in the range of 50-200 particles, where undercounting is significant yet somewhat predictable.

**GNR quantification at high densities.** Above 200 GNR per spot, accurate enumeration of GNRs becomes impossible due to severe crowding. However, we found that in this regime, the number of GNRs scales linearly with the total interferometric signal when integrated over the entire spot area (Figure 2e). This 'analog' measurement is only effective above about 200 GNR per spot because of variable background of other non-particle features, such as the boundary of the spot itself. Interestingly, on the highest concentration chips (10-100 pM) GNRs were packed sufficiently close together to lose their plasmonic scattering properties, and acted like a highly reflective gold film (Figure 3a). In those cases the total interferometric light scattering signal actually decreases, despite the spots becoming visible to the naked eye. We found we were able to robustly quantify the reflectivity of a spot simply by taking the average value of the z-stack at each $(x, y)$ position (rather than the maximal difference for interferometric detection, as in Figure 1c), and then normalizing by and subtracting the reflectivity of the film alone.

These three quantification methods—single GNR counting, total interferometric signal and total reflective signal—are complementary since they each quantify GNR binding within different ranges of concentration (Figure 3b). By combining them, the instrument achieves both single-nanoparticle readout sensitivity and a dynamic range of nearly one million, from a single scan.

Importantly, all three of these measurements are obtained from the same image data. Most other methods to achieve a large dynamic range require re-scanning the array multiple times, either after multiple rounds of silver enhancement[30], changing the objective lens, or changing the exposure time in the case of fluorescence.

To demonstrate the utility of this technique for large arrays, we incorporated a motorized stage and automation software into the instrument. The 10x objective provides a field of view of 1.46 mm$^2$, large enough to fit between 24 to 63 spots with a spot pitch of 250 μm or 150 μm, respectively. To scan larger arrays, the region of interest is divided into tiles that are sequentially scanned and then combined (Figure 4). Image acquisition takes about 30 seconds per field of view, and is primarily limited in speed by the amount of available light power (we used a 940 mW, 660 nm LED source with a 650nm, 10 nm FWHM band-pass filter)—a brighter illumination source could enable the same acquisition in as little as 5 seconds. With current settings, the instrument would be able to scan a 1cm$^2$ array in about 35 minutes, or a large array with 15,000 spots and a 150 μm pitch in about two hours (a brighter source could lower this to 5 and 20 minutes, respectively).

## CONCLUSIONS

Digital microarrays stand to improve the weakest performance aspects of microarrays (sensitivity and dynamic range) by about three orders of magnitude, while maintaining all the advantages of the microarray platform. Our approach combines gold nanoparticle labels with an interferometric detector to achieve single-molecule reporting capability with a 10x objective lens, allowing high-throughput acquisition of typical centimeter-scale microarrays.

We anticipate that this technique will have greatest utility in applications that require both high sensitivity and high multiplexing capability, such as mRNA and miRNA quantification. Since the sensitivity of digital microarrays may be increased by adding duplicate spots for each probe condition to increase the sensor surface area, this high-throughput method may also have utility even when the number of probes is few, but very high sensitivity is required. The instrument is no more complex or costly than commercial fluorescence readers, so it may be deployed in similar research or clinical diagnostic settings. In terms of assay compatibility and performance, this technique is fully compatible with the protocols in literature that use gold nanoparticle labels to detect RNA,[30] DNA[22], and proteins[15], and we expect it would have the similar limits of detection (from 100 attomolar to 1 femtomolar) since the biomolecular aspects are unchanged.

## METHODS

**Preparation of DNA microarrays on IRIS substrates:** IRIS chips were fabricated by performing 110 nm of thermal oxide growth, photolithographic patterning and oxide etching on polished silicon wafers (Silicon Valley Microelectronics, Santa Clara CA). Chips were coated with a co-polymer designed for high density immobilization of amine-terminated DNAs onto glass substrates (MCP-4, Lucidant Polymers, Sunnyvale CA). Amine-terminated DNA probes were purchased from Integrated DNA technologies (Coralville, IA) and printed onto the chips with a SCIENION, Inc S3 FlexArrayer following manufacturer instructions.

**Gold nanorod functionalization and microarray hybridization:** Citrate-stabilized GNRs were purchased from Nanopartz, Inc (Loveland, CO) and conjugated with the universal label sequences using the 'fast acid' protocol published by others.[31] The conjugated GNRs were washed and diluted to final concentrations in 1x (10 mM) phosphate buffered saline with 600 mM sodium chloride, 0.1% Tween-20, and 1mM EDTA. The IRIS microarray chips (1 cm$^2$ size) were placed individually into 24-well plate wells with 250 μL of the GNR containing solutions, and placed on an orbital shaker at room temperature for 4 hours. Chips were washed once in the hybridization buffer, then in phosphate buffer with 600 mM sodium (but without Tween-20), then finally with phosphate buffer with 150 mM sodium chloride before drying with nitrogen.

**Image acquisition and analysis:** The automated instrument was controlled using custom scripts and plugins for Micro-manager, an open-source microscope control application[32]. To acquire the interferometric image with the 10x objective, a z-stack of 15 frames with a 3 μm step size was acquired (with the 50x objective, 16 frames were acquired with a 250 nanometer step size).

# SUPPLEMENTAL FIGURES

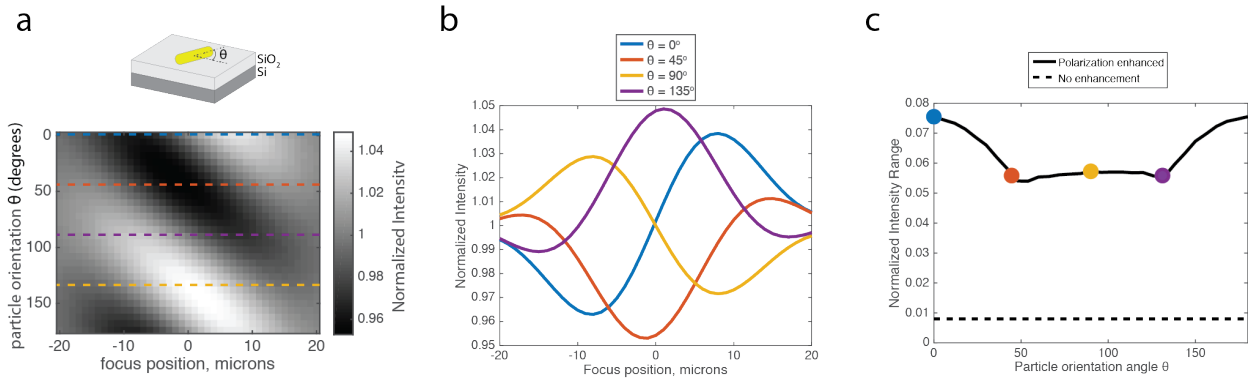

**Figure S1:** (a) Simulations of normalized intensity of plasmonic gold nanorods with the different surface orientation angle $\theta = [0^o, 180^o]$ at different focus positions, imaged with digital microarray instrument. (b) Four simulated cross sections from (a). Not every particle is visible at every focus position. For example, at at $z = 0$, scattered light from particles oriented at $0^0$ and $90^0$ are in quadrature with the reflected light, and they will not be visible in the image. (c) The 'normalized intensity range' of a particular particle is the difference between the maximum and minimum normalized intensity in the defocus profile of that particle (e.g., Figure 1c). Here, the normalized intensity range is simulated for all possible particle orientations in panel (a). Particles of all orientations are enhanced by between 5-fold and 7-fold, as compared with no polarization optics in the the collection path.

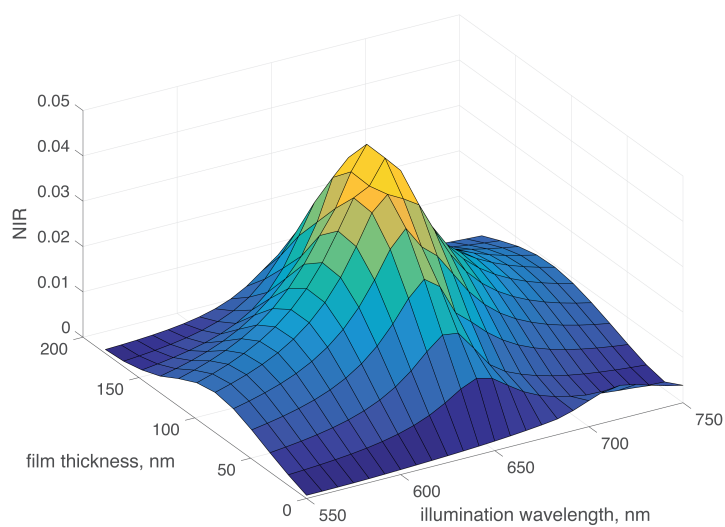

**Figure S2:** Simulation-based optimization of IRIS substrate oxide film thickness and illumination wavelength for plasmonic GNRs. For GNRs with a longitudinal surface plasmon resonance of about 660 nm, the optimal oxide thickness is about 110 nm. NIR: normalized intensity range, as in Figure S1c.

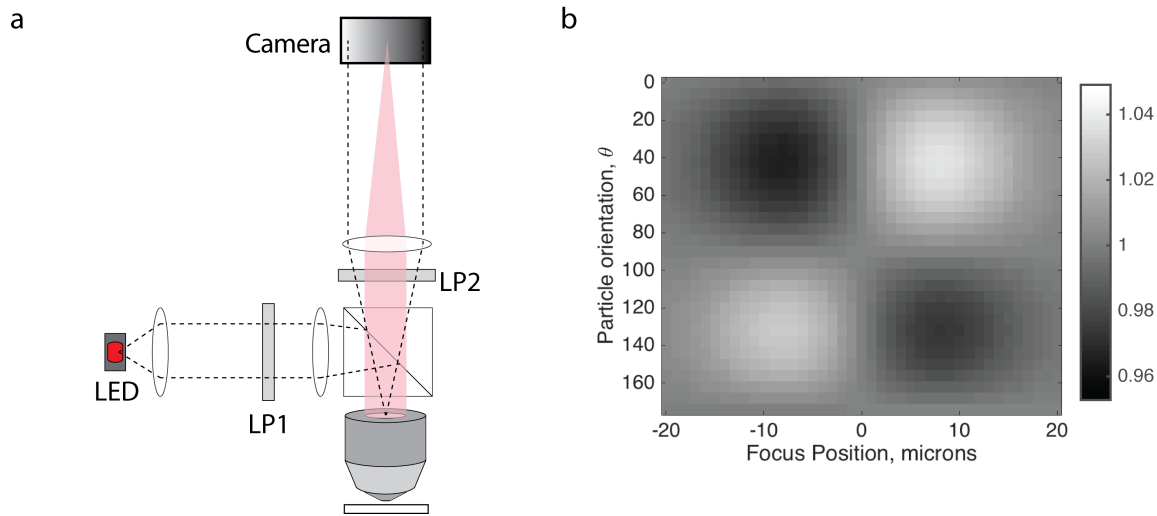

**Figure S3:** (a) Schematic of cross polarization approach for attenuating the reflected light (LP, linear polarizer). The polarizer in the collection path is adjusted to attenuate 95% of the reflected light. (b) Simulations of normalized intensity of GNRs with different surface orientation angle $\theta = [0^0, 180^o]$ at different focus positions, when imaged with the cross polarization approach—compare with Figure S1a. The illumination is linearly polarized along $\theta = 0°$, and the analyzer is oriented at 86°. Gold nanorods oriented at 0° and 90° are not visible at any focus position, since their light scattering is either highly attenuated by the analyzer (if $\theta = 0^o$) or the longitudinal surface plasmon is not excited (if $\theta = 90^o$).

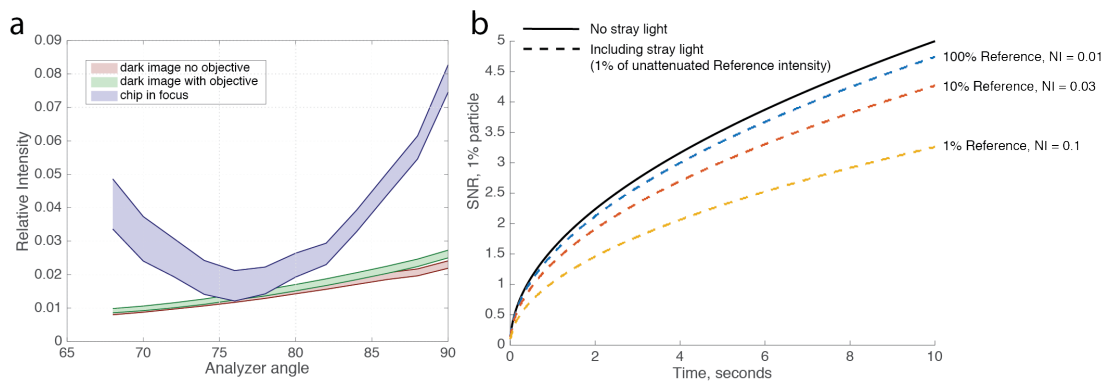

**Figure S4:** Experimental optimization of attenuation. (a) Relative intensity of the reflected light as the analyzer (LP2, Figure 1b) is rotated, compared to no analyzer. The intensity varies somewhat across the image: shaded regions indicate minimum and maximum values of the image histogram. To quantify the amount of stray light, dark images were taken by removing the IRIS chip and placing a piece of black felt cloth far from the focal plane. Removing the objective only slightly reduced the stray light (red vs green lines), indicating that the majority of the stray light comes from back reflections in the non-polarizing beam splitter cube. (b) Analysis of the effect of stray light shot noise on the signal to noise ratio of a particle, provided that the stray light is 1% the intensity of the reflected light. Attenuating the reference light increases the particle's normalized intensity (i.e. visibility), but will actually require longer to detect because the noise floor has increased due to the additional shot noise of the stray light.

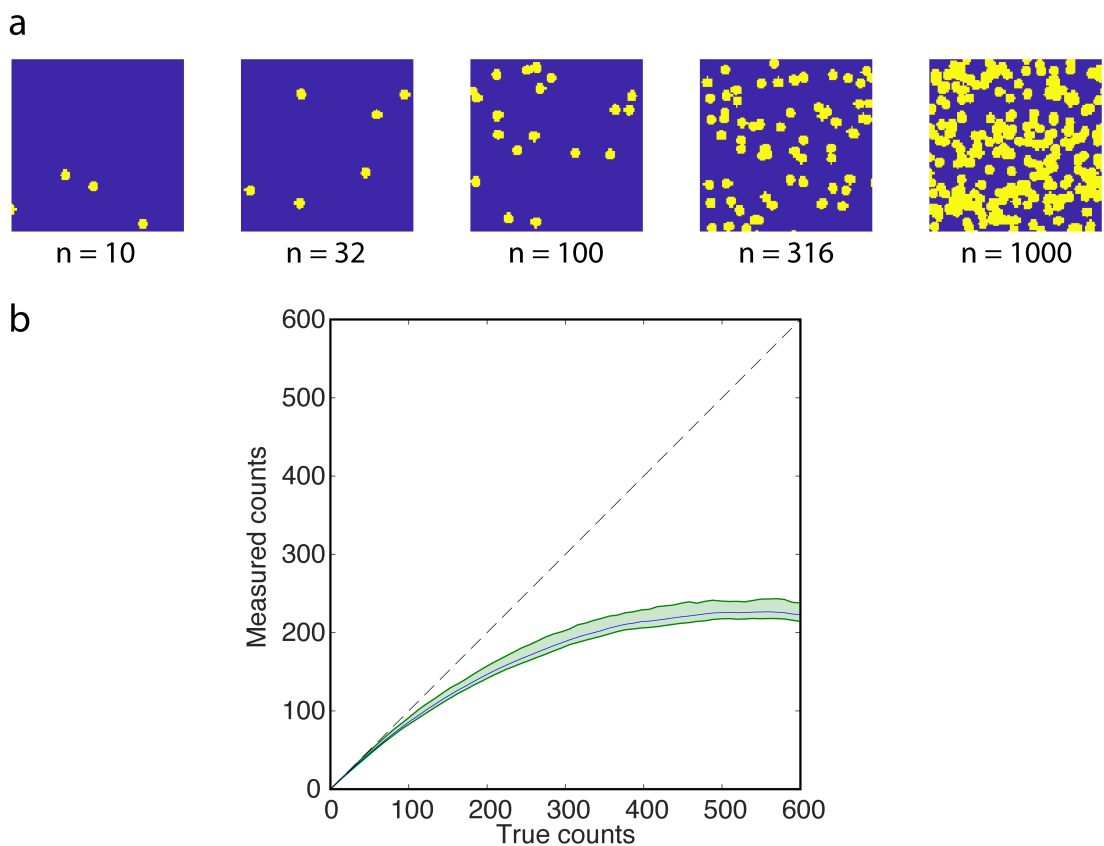

**Figure S5:** A simple model of particle under-counting due to crowding. Diffraction-limited images of GNRs were modeled as simple disks placed randomly within a region, and any overlapping disks are 'detected' as a single particle. (a) Regions with increasing numbers of disks (cropped to 25% to show detail). (b) Measured (i.e., 'detected') vs true number of particles for the simple model. Blue line indicates the ensemble average, and green lines indicate 10[th]-90[th] percentile range, for 1,500 total simulations. This approximately recapitulates the experimental observations in Figure 2d.